\documentclass[traditabstract]{aa}
\usepackage{graphicx}
\usepackage{natbib}
\usepackage{color}
\usepackage{txfonts}
\def\gr{$\gamma$-ray}

\begin{document}

\authorrunning{Neronov et al.}
\titlerunning{No evidence for \gr\ halos}

   \title{No evidence for gamma-ray halos around active
  galactic nuclei resulting from intergalactic magnetic
  fields}

   \author{A.~Neronov\inst{1,2}
          \and
          D.V.~Semikoz\inst{3,4}
        \and P.G. Tinyakov\inst{4,5}
          \and I.I. Tkachev\inst{4}
          }

   \institute{ISDC Data Centre for Astrophysics, Ch. d'Ecogia 16, 1290 Versoix, Switzerland
  \and Geneva Observatory, Ch. des Maillettes 51, 1290 Sauverny, Switzerland     
  \and  APC, 10 rue Alice Domon et Leonie Duquet, F-75205 Paris Cedex 13, France
  \and Institute for Nuclear Research RAS, 60th October Anniversary prosp. 7a,
Moscow, 117312, Russia
\and Service de Physique Th«eorique,  Universit«e Librede Bruxelles,
CP225, blv. du Triomphe, B-1050, Bruxelles, Belgium}
   \date{Received ; accepted }

\abstract
{We analyze the gamma-ray halo around stacked AGNs reported in
Ap.J.Lett., 2010, 722, L39. First, we show that the angular distribution of
$\gamma$-rays around the stacked AGNs is consistent with the angular
distribution of the $\gamma$-rays around the Crab pulsar, which is a point source
for {\it Fermi}/LAT. This makes it unlikely that the halo is caused by
  an electromagnetic cascade of TeV photons in
the intergalactic space. We then compare the angular distribution of
$\gamma$-rays around the stacked AGNs with the point-spread function
(PSF) of {\it Fermi}/LAT and confirm the existence of an excess above
the PSF. However, we demonstrate that the magnitude
and the angular size of this effect is different for photons converted
in the front and back parts of the Fermi/LAT instrument, and thus is
an instrumental effect. }


\keywords{gamma rays: galaxies Ð galaxies: active Ð magnetic fields}
   \maketitle

High-energy gamma rays propagating through the Universe produce
electron-positron pairs in interactions with the extragalactic
background light \citep{kneiske04,franceschini08,stecker09}. The leading particle of the pair then upscatters the
background photons through the inverse Compton effect, creating an
electromagnetic cascade. If there is a weak intergalactic
magnetic field, electrons and positrons in the cascade are deflected
from the original direction before being converted back to
photons. These secondary photons create a "halo" of softer gamma-ray photons around an
extragalactic TeV gamma-ray source \citep{coppi,plaga,neronov07,japanese,elyiv09,kachelriess09,kachelries}. This process may be used to
constrain the parameters of the extragalactic magnetic fields \citep{plaga,neronov07,japanese,elyiv09}. 
The absorption of TeV \gr s leads to the cascade emission in the GeV energy range, which is accessible for observations with the {\it Fermi} telescope \citep{neronov09}.  Non-detection of extended emission around blazars with hard intrinsic spectra extending to the multi-TeV energy range by {\it Fermi} was used to derive a lower boundary on magnetic fields in the intergalactic medium at the level of $10^{-17}$~G to $10^{-15}$~G, depending on the assumptions about the intrinsic spectral properties of the analyzed sources  \citep{neronov10,tavecchio10a,tavecchio10b,dolag10}.

\citet{Ando:2010rb} recently claimed to posess evidence for the gamma-ray halos
around AGNs from the {\it Fermi} source catalog. The halo was found in
the stacked signal ofAGNs selected by certain criteria. It was
interpreted to be caused by cascading and deflections of TeV photons
in the extragalactic magnetic fields.  Unfortunately, we find that the proposed
interpretation of the halo is unlikely, but a more probable cause is a
tail in the point-spread function (PSF) of photons pair converted in the back thick layer of the {\it Fermi} detector.

\begin{figure*}
\begin{center}
\includegraphics[width=0.49\linewidth]{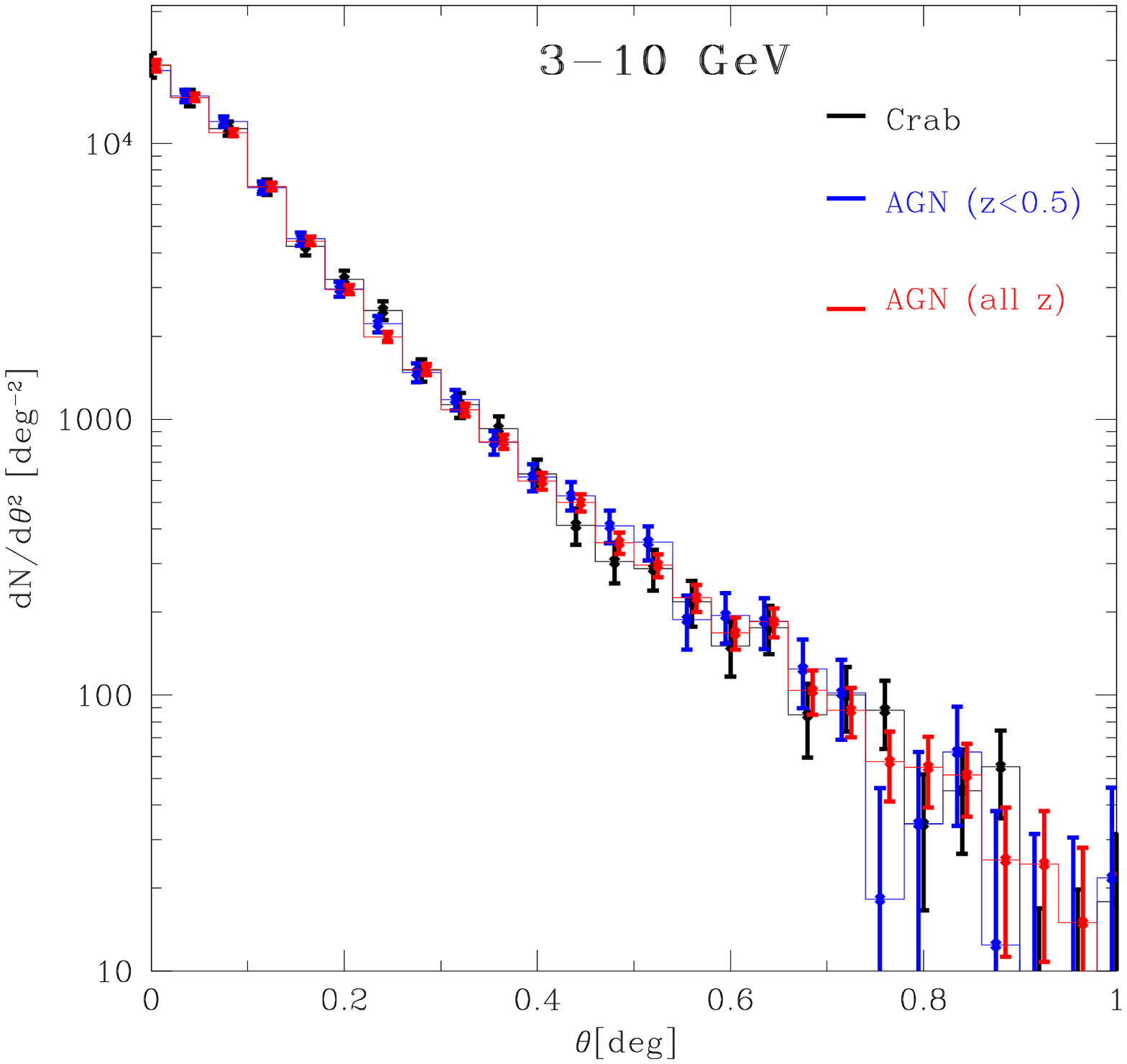}
\includegraphics[width=0.49\linewidth]{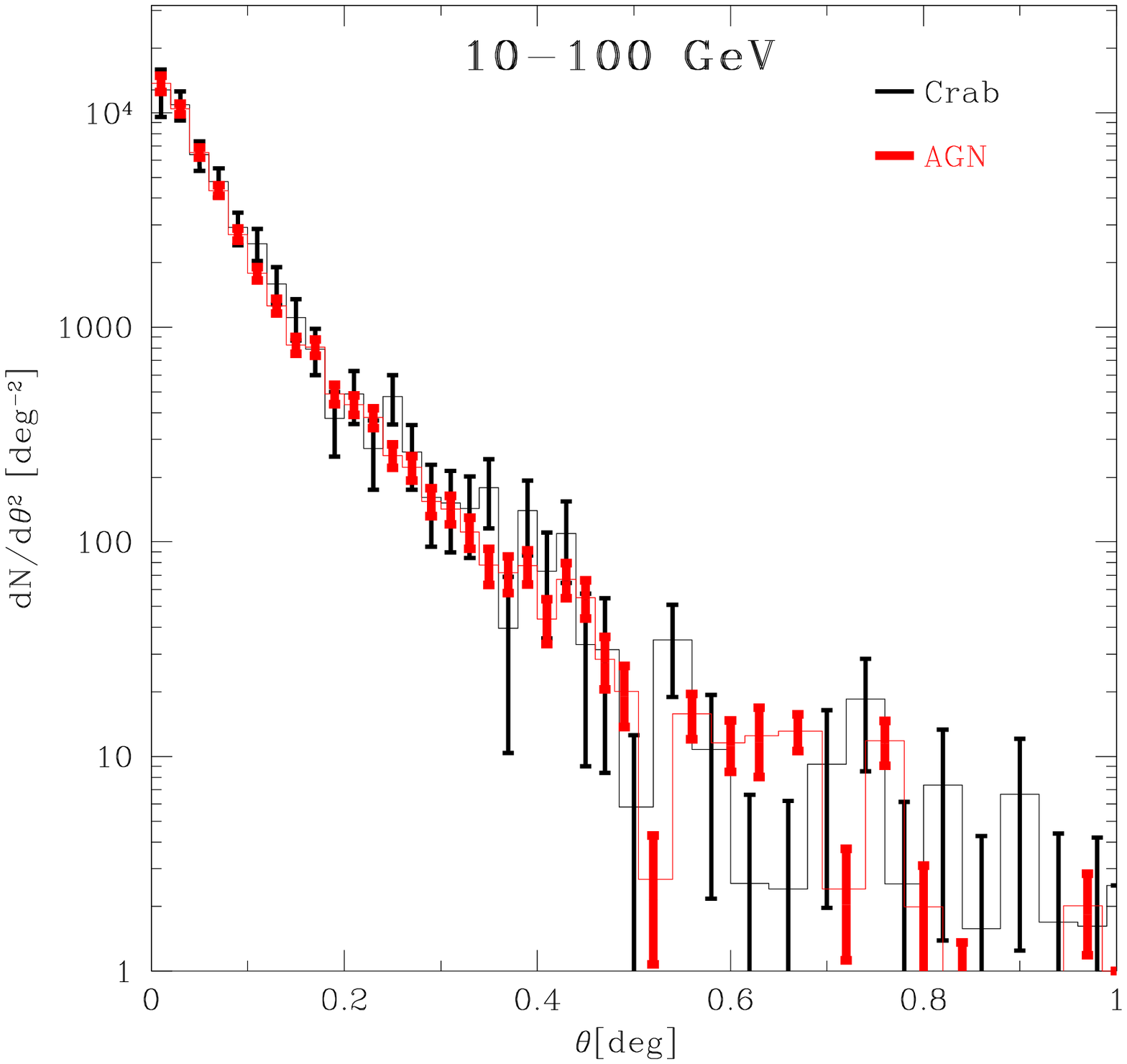}
\end{center}
\caption{Comparison of the background-subtracted angular distributions of $3-10$~GeV (left) $10-100$~GeV (right)
  $\gamma$-rays around the position of the Crab pulsar (black) and
  around the stacked AGNs (red).  Normalization of the distribution of  $\gamma$-rays around AGN is scaled to match the $\gamma$-ray
  distribution around the Crab pulsar. In the 3-10 GeV band AGN with redshift $z<0.5$ are shown in blue. }
\label{fig:CRAB_vs_AGN}
\end{figure*}

Our argument consists of two parts. 

First, we compare the stacked AGN signal  to the signal of the Crab pulsar. This is a bright
galactic gamma-ray source whose signal in the {\it Fermi} energy band
consists of two contributions: emission from the pulsar and from the
associated pulsar wind nebula (PWN). The Crab PWN has an angular size $\simeq
0.05^\circ$ \citep{Hester:2008}, which is below the angular resolution
of the LAT telescope onboard {\it Fermi}. This means that the Crab PWN is
a point source for LAT. In addition, this PWN is situated in the Galactic
anti-center region, where the density of sources in the multi-GeV
energy band is low and the diffuse Galactic $\gamma$-ray background is
relatively low as well. This means that Crab signal is not
contaminated by the nearby point sources or by strong
inhomogeneities of the diffuse background, and the source can be
considered as an isolated point source.

The LAT PSF depends on the photon energy. Thus, the shape of the
angular distribution of $\gamma$-rays around a source depends on the
source spectrum.  The total Crab pulsar+PWN spectrum in the 10-100 GeV
band is well described by a power law $dN/dE\sim E^{-\Gamma}$ with the
photon index $\Gamma\simeq 2$ \citep{Abdo:2009ec}, which is the same as
the cumulative spectrum of the AGNs derived by \citet{Ando:2010rb}.  This implies that the point-source
contribution to the AGN signal should have the same angular shape as
the Crab signal. Thus, a difference between the angular
distributions of photons around AGN and the Crab PWN would indicate a halo around AGNs in addition to the point-source
contribution. 

Fig.~\ref{fig:CRAB_vs_AGN} shows the angular distribution of photons
around the stacked AGNs (red) and the Crab source (black). In both cases the
background is subtracted. The shapes of the two signals coincide,
which means that the entire stacked AGN signal is well described by a
point-source signal, with no additional halo contribution. In an update of their analysis, \citet{Ando:2010rb} have claimed that the halo signal appears only in a subset of AGN at low redshift, $z<0.5$ and only in the 3-10 GeV energy band. To allow an explicit comparison with Fig. 4 of   \citet{Ando:2010rb},  we show in blue the photon distribution around  AGN at $z<0.5$ in the left panel of Fig.  \ref{fig:CRAB_vs_AGN}, which corresponds to photon distributions in the 3-10 GeV band. No discrepancy between the Crab PWN and AGN profiles is seen in this case either.

Second, we have investigated the nature of the excess in the angular 
distribution of photons around AGN above the LAT PSF in the 
energy range 10-100 GeV.
$\gamma$-rays detected by the LAT telescope are split into two types: photons   that are pair-converted in the thin front layer of the LAT
detector ("front" photons), and photons converted in the thick back
layer ("back" photons) \citep{Abdo:2009gy}. The shapes of the PSF for these two types of photons are significantly different. Taking this into account, we split the entire photon signal from
the set of AGNs considered by \citet{Ando:2010rb} into two parts corresponding to the front and back converted photons and analyze each part  separately.   We then compare the angular distributions of the
front and back photons with the corresponding PSFs calculated using the {\tt
  DIFFUSE\_P6\_v3} calibration files by averaging the PSF for a given
photon energy and incidence angle over the entire set of energies and
incidence angles of detected photons from the AGN set. 

\begin{figure*}
\begin{center}
\includegraphics[width=0.49\linewidth]{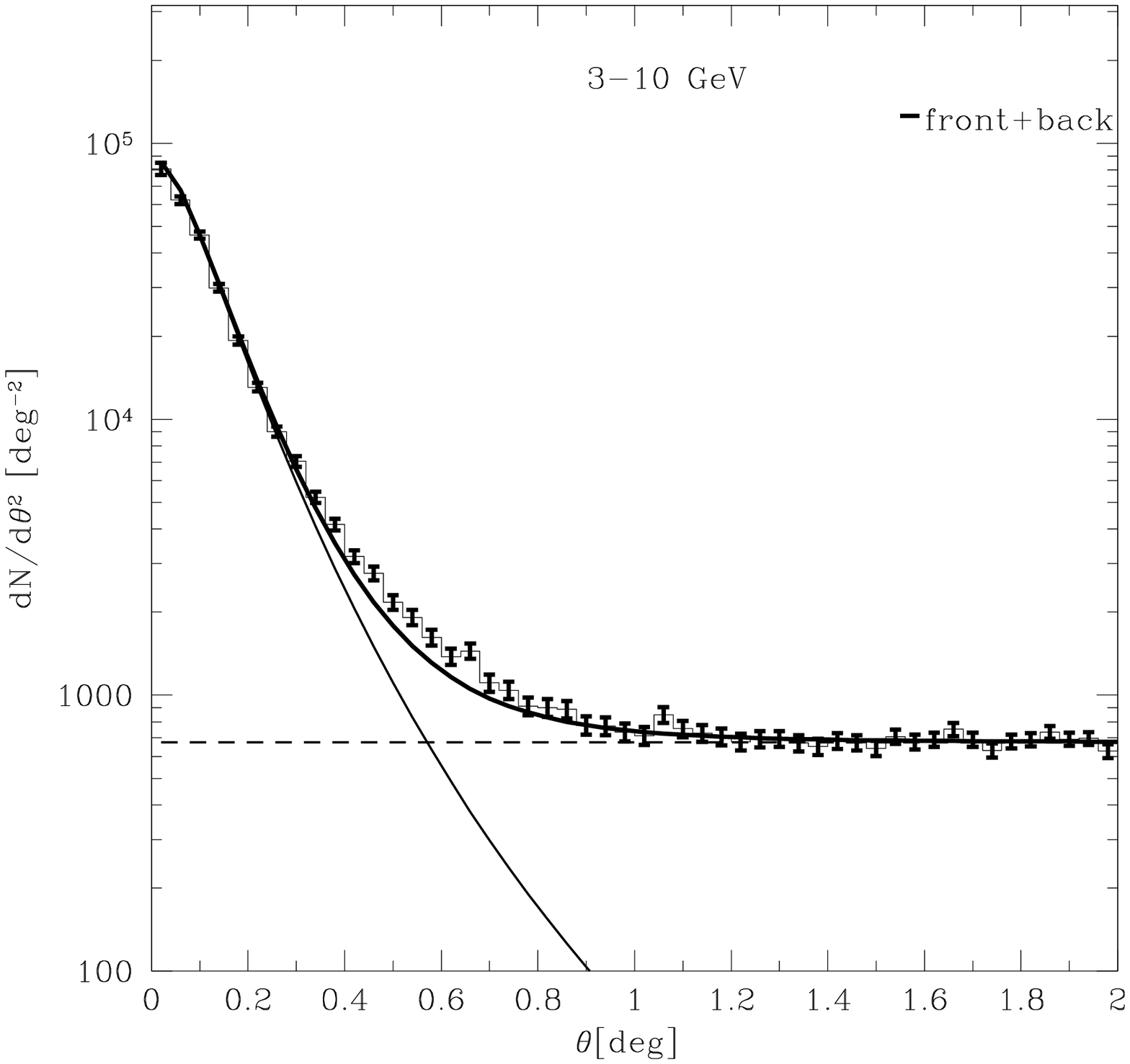}
\includegraphics[width=0.49\linewidth]{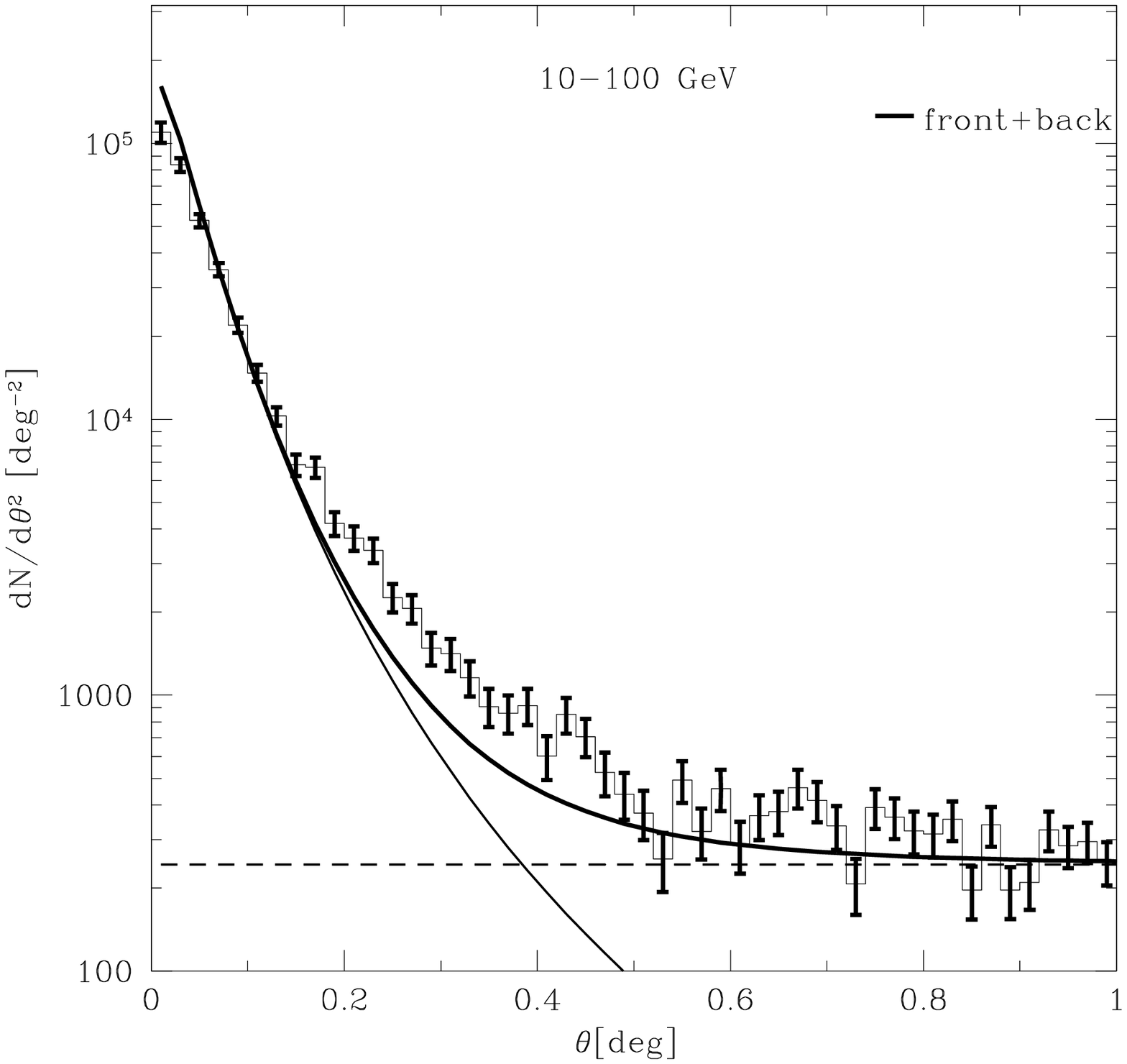}
\end{center}
\caption{Angular distribution of photons around stacked AGNs. Curves
  show the PSF (thin solid line), the background level (dashed line)
  and PSF plus background (thick solid line).  }
\label{fig:AGN_vs_PSF}
\end{figure*}

The angular distribution of all (front + back) photons around AGN is
shown in Fig.~\ref{fig:AGN_vs_PSF} together with the corresponding
PSF. We find an excess in the data at $0.2^\circ<\theta<0.9^\circ$ for the 10-100 GeV band and in the $0.5^\circ-1^\circ$ range in 3-10GeV band. There is a significant difference between our
estimate of the background level shown in Fig.~\ref{fig:AGN_vs_PSF} and the background level in the Fig.~2 of \citet{Ando:2010rb}, which is a factor of $\simeq
3$ higher than our estimate in the 10-100 GeV band. 
The constant background level shown in Fig. \ref{fig:AGN_vs_PSF} was derived from fitting of the data in the range $0<\theta<2^\circ$ around AGN with a model consisting of PSF plus a constant.

We have checked that our estimate of the background level is
self-consistent in the following way. The total number of photons
detected by {\it Fermi} in the energy band 10-100 GeV in the Galactic
latitude range $|b|>10^\circ$ is $N_{\rm total}\simeq 5.6\times
10^4$. Only a small fraction of these photons, $N_{\rm source}\simeq
0.6\times 10^4$ could be associated to the known {\it Fermi}
sources. The remaining photons contribute to the diffuse Galactic and
extragalactic backgrounds.  Dividing the number of diffuse background
photons by the solid angle $\Omega$ spanned by the considered part of
the sky, one finds the surface brightness of the background
$dN/d\theta^2\simeq (N_{\rm total}-N_{\rm source})/\Omega\simeq
1.5$~deg$^{-2}$. Multiplying this number by the number of AGN
considered in the analysis one arrives at the background estimate
consistent with the one shown in Fig. \ref{fig:AGN_vs_PSF}. We stress
that this background estimate is significantly lower than that found by
\citet{Ando:2010rb}

\begin{figure*}
\includegraphics[width=0.49\linewidth]{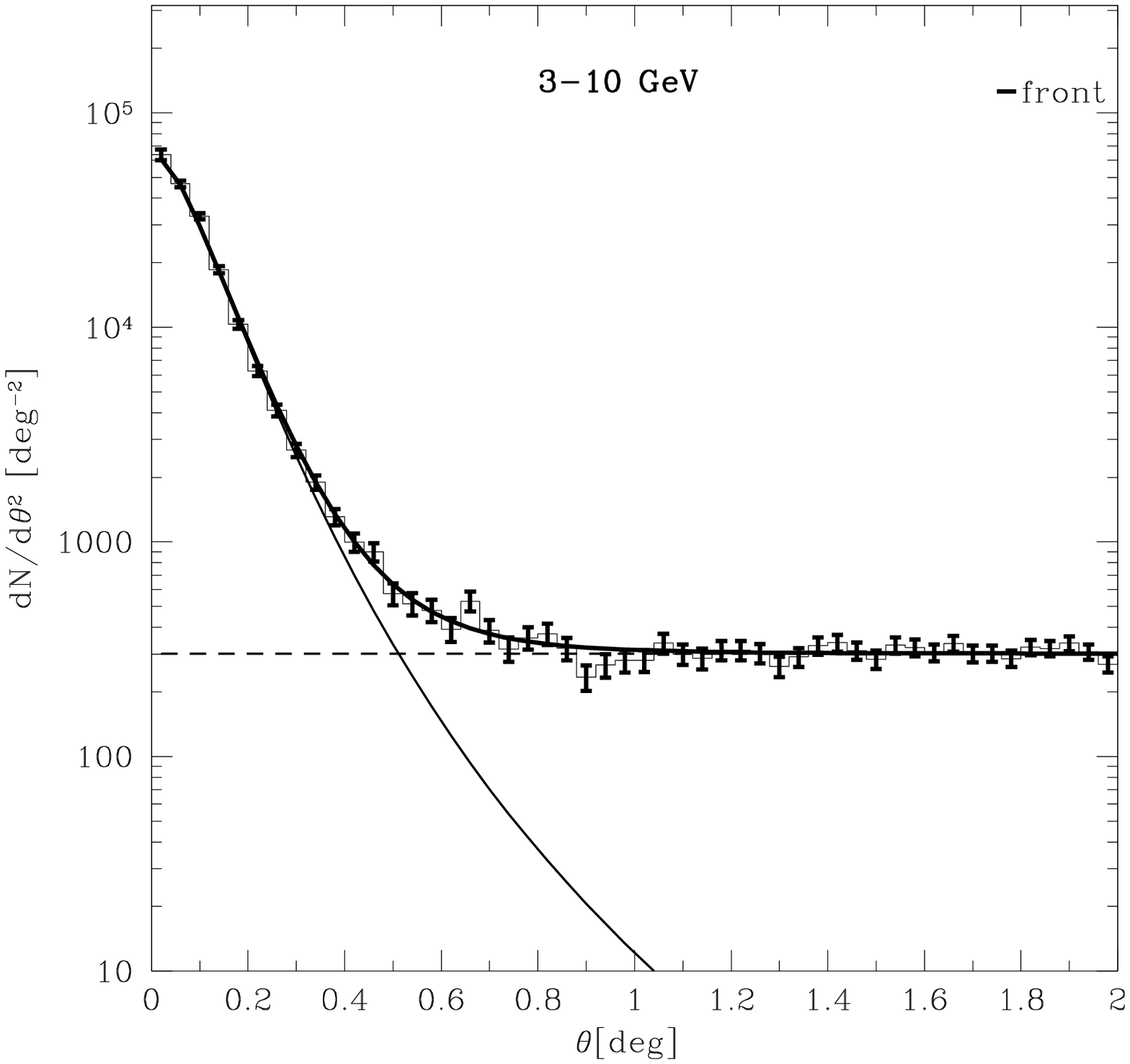}
\includegraphics[width=0.49\linewidth]{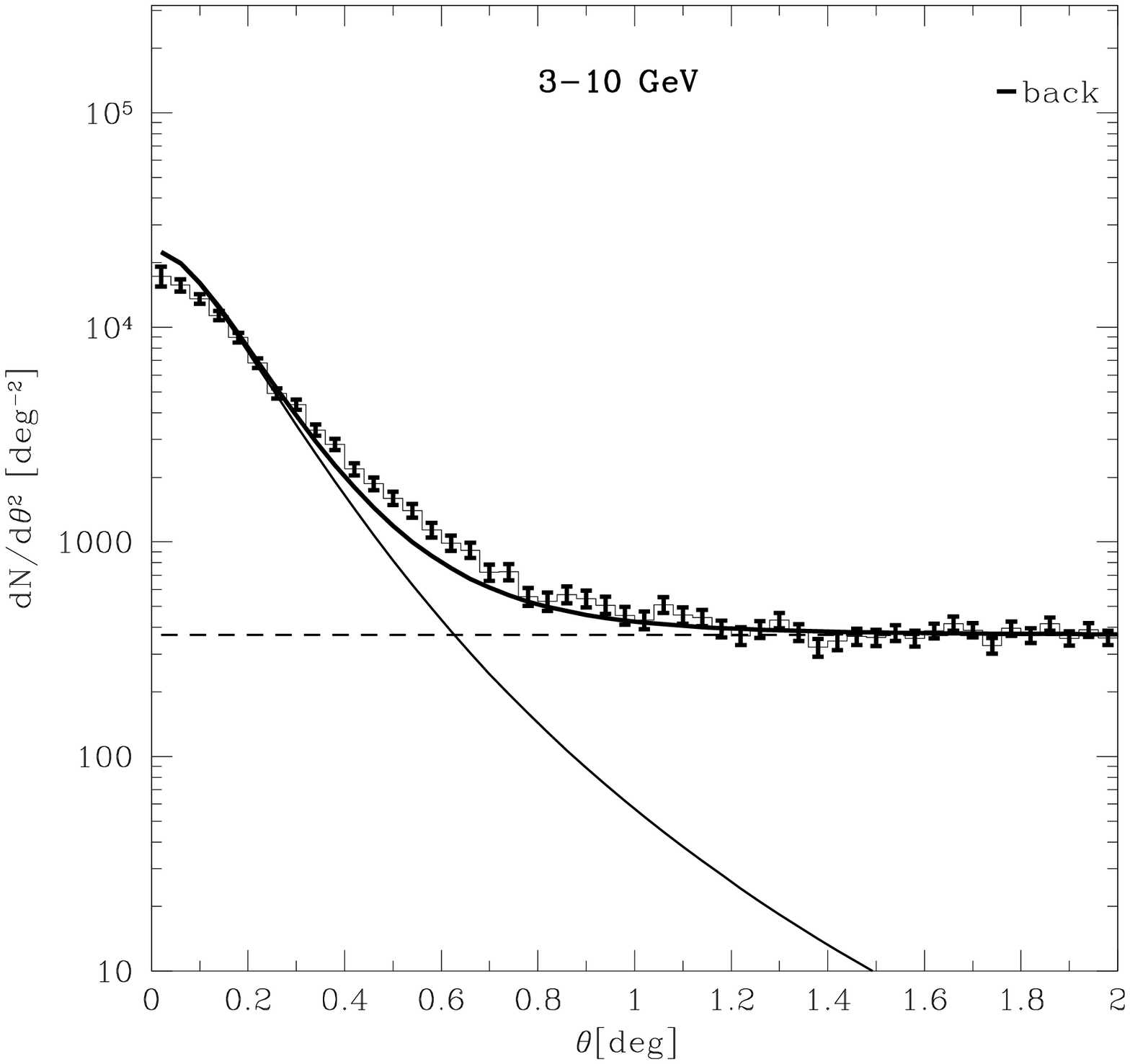}
\includegraphics[width=0.49\linewidth]{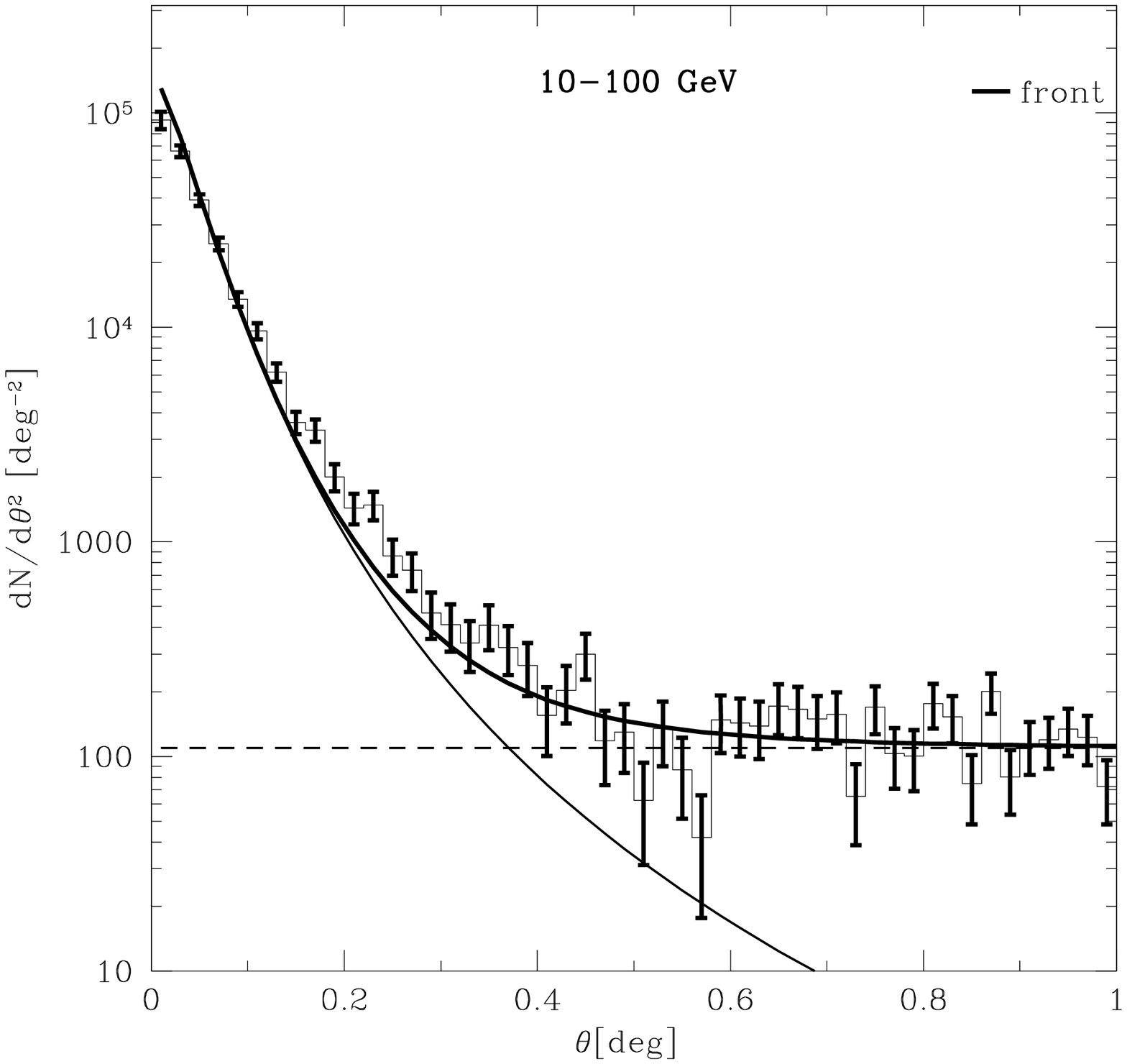}
\includegraphics[width=0.49\linewidth]{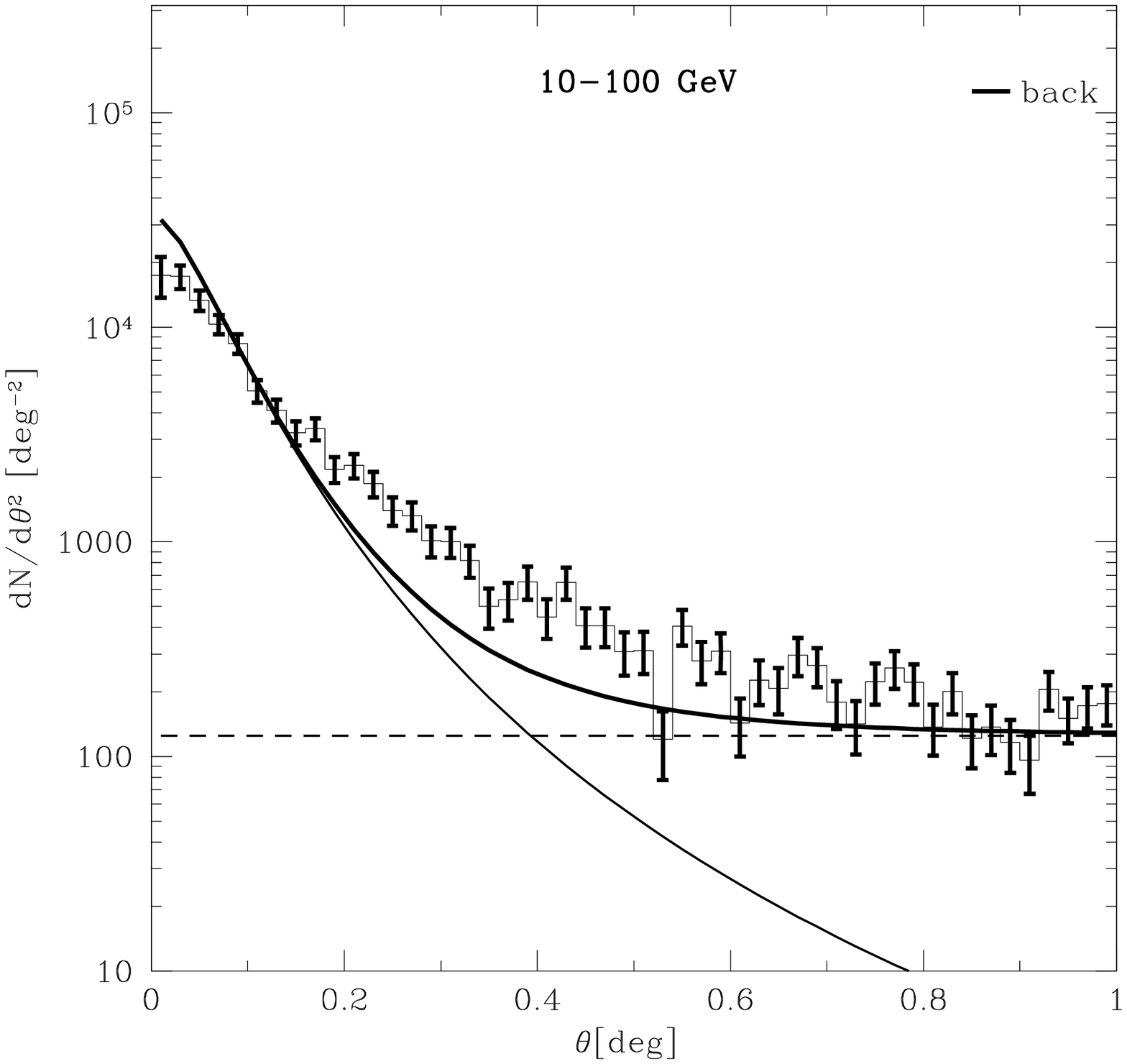}
\caption{The same as Fig.2, but separately for the front (left panel) 
and back (right panel) photons in 3-10 GeV band (top) and 10-100 GeV band (bottom). }
\label{fig:AGN_vs_PSF-front_back}
\end{figure*}

The angular distributions of front and back photons separately are
shown in Fig.~\ref{fig:AGN_vs_PSF-front_back}, each with the
corresponding PSF. One can see that the front photons have more
compact PSF, which is reasonably compatible with the angular
distribution of the front photon part of the AGN signal. The fit of
the data in the range $0<\theta<2^\circ$ by the PSF + background gives $\chi^2/{\rm dof} = 125/98$
($p\sim 3\%$) in the 10-100 GeV band. Top panels of Fig. 3 show the angular distributions of front and back photons in the 3-10 GeV band. Obviously there is no hint of excess above the PSF of the front photons in this energy band. The reduced $\chi^2$ of the fit is $\chi^2/{\rm dof}=67/72$.

The back photons have a wider PSF, which does not reproduce the back part
of the AGN signal in either 3-10 GeV or 10-100 GeV energy range. 

Adding front and back photon signals one can verify that most of the excess above the {\tt DIFFUSE\_P6\_v3} PSF in Fig \ref{fig:AGN_vs_PSF} is caused by the back photons. There is no strong excess above the PSF in the front photon signal. This indicates that the excess of the photons above the PSF, reported in \citet{Ando:2010rb}, cannot be attributed to the extended emission around AGN, because in this case an equally strong excess should be detected in both the front and back photon signals.

\begin{figure*}
\includegraphics[width=0.49\linewidth]{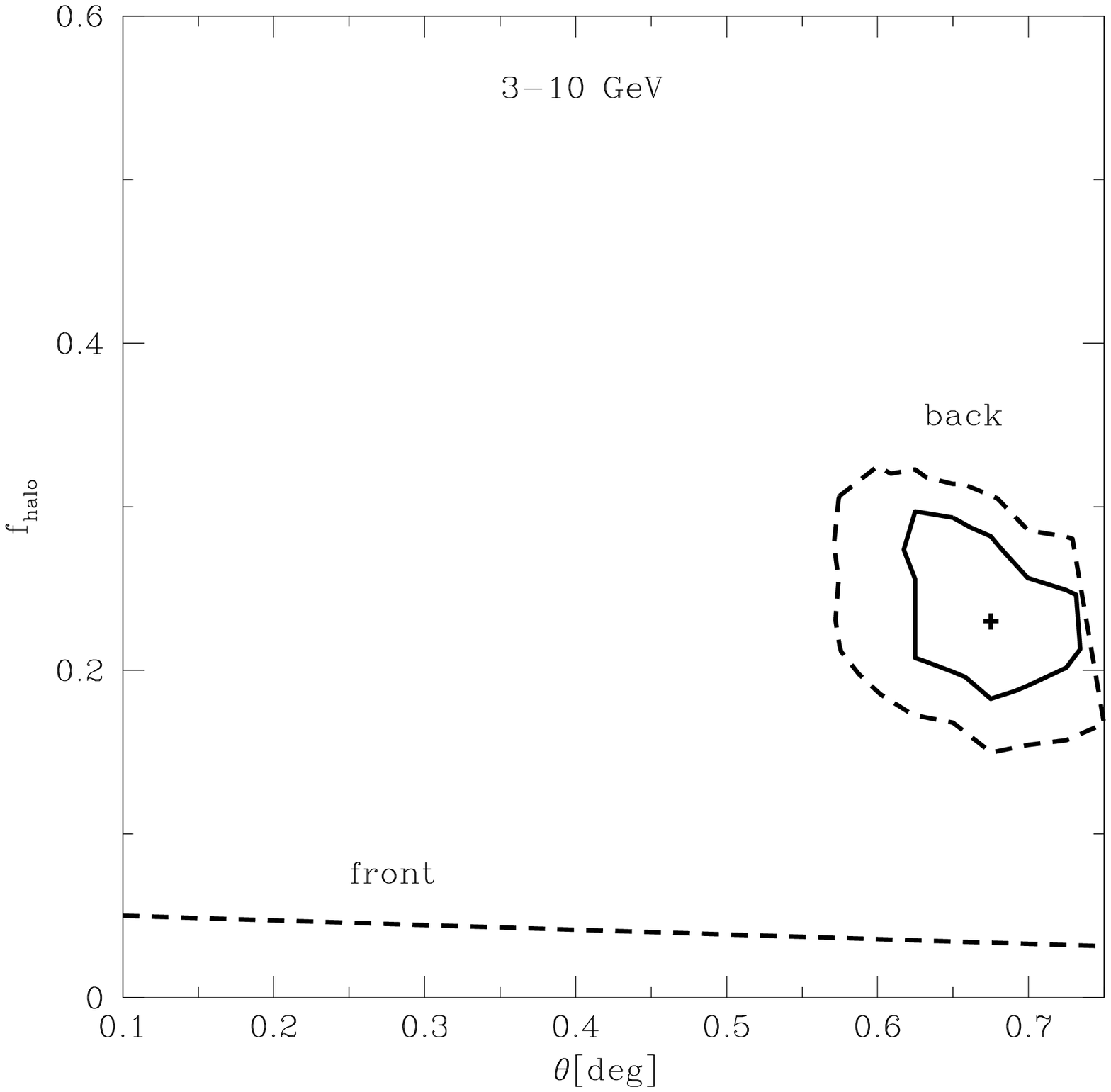}
\includegraphics[width=0.49\linewidth]{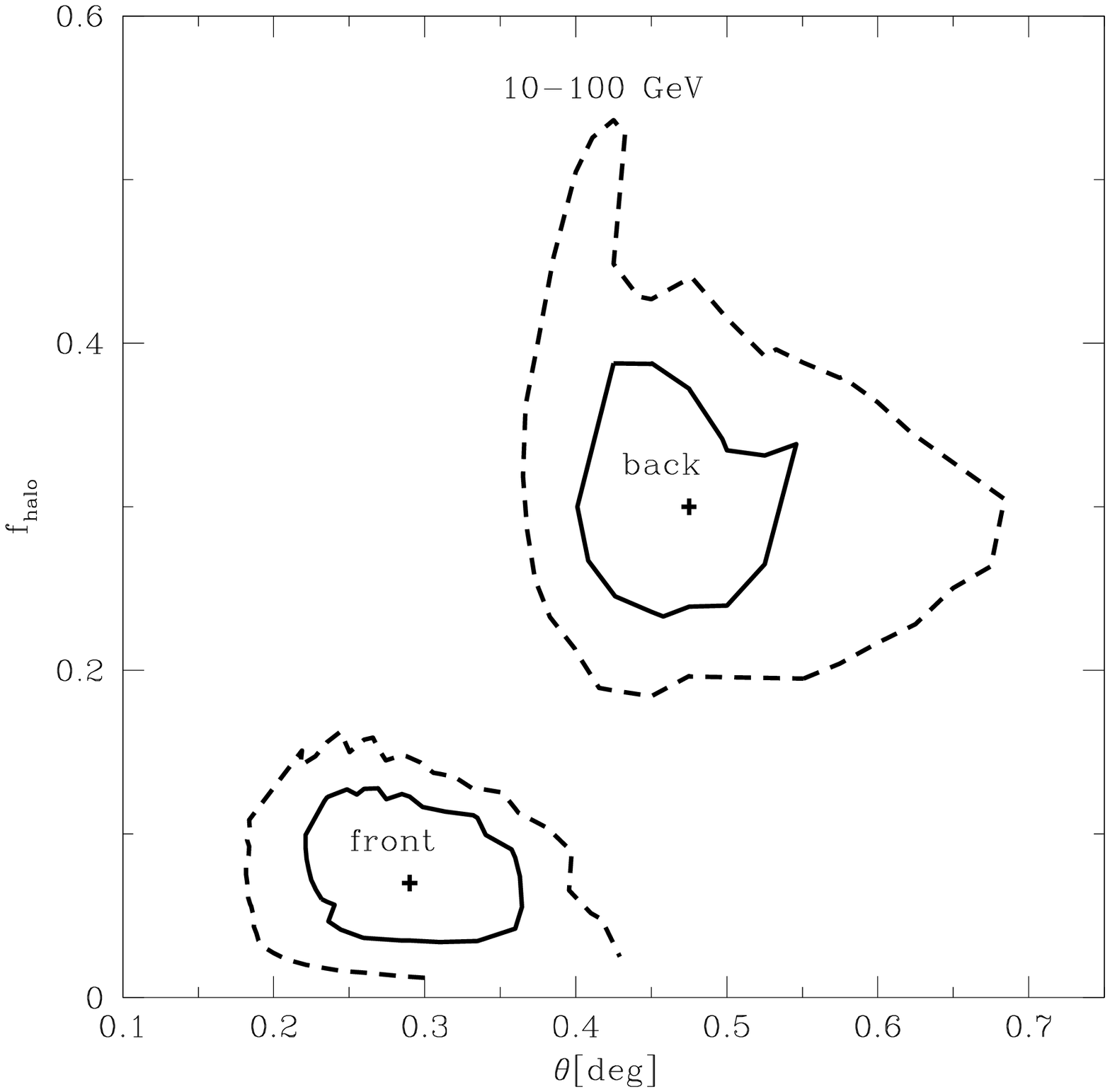}
\caption{68 \% (solid) and 99\% (dashed) confidence contours for the halo parameters $\theta_{\rm halo}, f_{\rm halo}$ derived from the separate analysis of front and back photons.}
\label{fig:confidence}
\end{figure*}

To quantify the excess in the front and back photon signals above the
{\tt DIFFUSE\_P6\_v3} PSF we followed \citet{Ando:2010rb} and introduced an additional $\theta^2$-Gaussian component $dN/d\theta^2\sim \exp\left(-[\theta/\theta_{\rm halo}]^4\right)$ in the model of the angular photon distribution. This additional
component depends on two parameters, $f_{\rm halo}$  (fraction of the
source signal contained in the additional component) and  $\theta_{\rm
  halo}$ (angular size of the core).  Fitting the angular photon
distributions with the improved model we found the best-fit
values and confidence regions for $f_{\rm halo}, \theta_{\rm halo}$
separately for the front and back photons. The result is shown in
Fig. \ref{fig:confidence}. The parameters of the additional components found
for the front and back photons are clearly different. An addition of the new component to the model does not improve the fit of the front photon angular distribution in the 3-10 GeV. Only an upper bound on the flux in the "halo" component could be derived, as is shown in the left panel of Fig. \ref{fig:confidence}. This confirms
our conclusion that the excess above the PSF is not caused by the real extended emission around AGN.

Indeed, the size and the flux of a real halo around AGN cannot depend on the way the photons were converted into $e^+e^-$ pairs in the {\it Fermi}/LAT detector. Instead, the excess above the   PSF in the back photons should be attributed to the imperfect modeling of the real PSF for back photons in the {\tt DIFFUSE\_P6\_v3} version of the LAT instrument characteristics. 
 
To summarize, the angular distribution of photons around the stacked
AGNs matches well that around the Crab source in both the $3-10$~GeV and
$10-100$~GeV bands. A separate analysis of the front and back converted photons shows that there is no evidence for a physical halo around AGN, neither in 3-10 GeV nor in 10-100 GeV band.  No halo caused by the extragalactic magnetic
fields is observed. 

{\it Acknowledgements.} We thank A.Kusenko and S.Ando for the discussion of the subject. The work of PT  is supported in part by IISN, Belgian Science Policy (under
contract IAP V/27). The work of AN is supported by the Swiss National Science Foundation grant PP00P2\_123426 .

\end{document}